\documentclass[conference,a4paper]{IEEEtran}
\usepackage[left=1.43cm,right=1.43cm,top=1.92cm,bottom=4.5cm]{geometry}

\usepackage[algo2e]{algorithm2e} 
\usepackage{amsmath,amssymb,amsmath,amsfonts}
\usepackage{bm}
\usepackage{bbm}
\usepackage{graphicx}
\usepackage{cite}
\usepackage{epstopdf}
\usepackage{gensymb}
\usepackage{color}
\usepackage{lipsum}
\usepackage{float}
\usepackage{epstopdf}
\usepackage[mathcal]{eucal}
\usepackage{subfiles}
\usepackage[T1]{fontenc}
\usepackage{siunitx}
\usepackage{tabulary}
\usepackage{epsfig,graphics,graphicx,subfig,amssymb,amstext,amsmath,algorithm,algorithmic,wrapfig,multirow}
\usepackage{array}
\usepackage{adjustbox}
\usepackage[dvipsnames]{xcolor}
\usepackage{cite}
\usepackage{times}
\usepackage{placeins}
\usepackage{textcomp}
\usepackage[font=small]{caption}
\usepackage{flushend}
\usepackage{color, colortbl}
\usepackage{tabularx}
\usepackage{bm}
\usepackage{enumerate}
\usepackage{tikz}
\usepackage{pgfplots}
\usepackage{epstopdf}
\usetikzlibrary{shapes,arrows}
\usepackage[utf8]{inputenc}
\usepackage{mathtools}
\usepackage[utf8]{inputenc}
\usepackage{xcolor}
\usepackage{tikz}
\usetikzlibrary{chains,arrows,calc,positioning}
\usepackage{amssymb}% http://ctan.org/pkg/amssymb
\usepackage{pifont}% http://ctan.org/pkg/pifont
\usepackage{soul}
\usepackage{breqn} % to split equations using dmath env
\usepackage{booktabs} % nice rules in tables
\usepackage{multirow}
\usepackage{enumitem}
\usepackage{tabulary}
\usepackage[normalem]{ulem}
\usepackage{dblfloatfix}
\usepackage{makecell}
\usepackage{lipsum}
\usepackage{amsthm}
\usepackage{comment}
\usepackage{filecontents}
\newtheoremstyle{mystyle}%                % Name
  {}%                                     % Space above
  {}%                                     % Space below
  {\itshape}%                             % Body font
  {}%                                     % Indent amount
  {\bfseries}%                            % Theorem head font
  {.}%                                    % Punctuation after theorem head
  { }%                                    % Space after theorem head, ' ', or \newline
  {}%                                     % Theorem head spec (can be left empty, meaning `normal')
\theoremstyle{mystyle}

\newlength \figwidth
\definecolor{bittersweet}{rgb}{1.0, 0.44, 0.37}
\definecolor{glaucous}{rgb}{0.38, 0.51, 0.71}
\definecolor{gainsboro}{rgb}{0.86, 0.86, 0.86}
\definecolor{babyblueeyes}{rgb}{0.63, 0.79, 0.95}
\definecolor{silver}{rgb}{0.75, 0.75, 0.75}
\definecolor{neoncarrot}{rgb}{1.0, 0.64, 0.26}
\definecolor{Gray}{gray}{0.9}
\definecolor{LightCyan}{rgb}{0.88,1,1}
\definecolor{BackgroundLightBlue}{rgb}{0.97,0.97,1}
\definecolor{BackgroundGray}{gray}{0.98}

\newcommand{\blue}[1]{\textcolor{blue}{#1}}

\newcommand{\green}[1]{{\textcolor[rgb]{0,0.5,0}{#1}}}

\makeatletter

 \let\oldforeign@language\foreign@language
 \DeclareRobustCommand{\foreign@language}[1]{%
   \lowercase{\oldforeign@language{#1}}}

\def\nb0{{\mathbf{0}}}
\def\nb1{{\mathbf{1}}}

\def\ncalA{{\mathcal{A}}}

\def\ncalG{{\mathcal{G}}}

\def\ncalR{{\mathcal{R}}}

\def\ncalU{{\mathcal{U}}}

\def\sinr{\mathtt{SINR}}			% Signal to interference plus noise ratio

\def\calB{\mathcal{B}}

\makeatother

\IEEEoverridecommandlockouts
\begin{document}

% This code is to reduce the list of authors by using et. al:
\bstctlcite{IEEEexample:BSTcontrol}

%\title{Large-scale Cellular Network Design via Data-driven Optimization and Transfer Learning}

\title{Data-driven Optimization and Transfer Learning for Cellular Network Antenna Configurations}

\author{\IEEEauthorblockN{Mohamed Benzaghta$^{\star}$, Giovanni Geraci$^{\dagger\,\star}$, David L\'{o}pez-P\'{e}rez$^{\sharp}$, and Alvaro Valcarce$^{\flat}$ \vspace{0.1cm}
}
\\ \vspace{-0.3cm}
\normalsize\IEEEauthorblockA{$^{\star}$\emph{Univ. Pompeu Fabra, Barcelona, Spain} \enspace \enspace $^{\dagger}$\emph{Telefónica Research, Barcelona, Spain}  \\ $^{\sharp}$\emph{Univ. Politècnica de València, Spain} \enspace \enspace  $^{\flat}$\emph{Nokia Bell Labs, Massy, France}}

%\thanks{This work was in part supported by the Spanish Research Agency through grants PID2021-123999OB-I00, CEX2021-001195-M, and CNS2023-145384, by the UPF-Fractus Chair, by the Spanish Ministry of Economic Affairs and Digital Transformation and the European Union NextGenerationEU through the UNICO 5G I+D SORUS project, and by Generalitat Valenciana through the grant CIDEGENT PlaGenT, CIDEXG/2022/17, Project iTENTE.}

\thanks{
This work was supported by 
\emph{a)} the Spanish State Research Agency through grants PID2021-123999OB-I00 and CEX2021-001195-M, 
\emph{b)} the UPF-Fractus Chair on Tech Transfer and 6G, 
\emph{c)} the Spanish Ministry of Economic Affairs and Digital Transformation and the European Union NextGenerationEU through actions CNS2023-145384, CNS2023-144333, and the UNICO 5G I+D SORUS project, \emph{d)} the Generalitat Valenciana, Spain, through the  CIDEGENT PlaGenT, Grant CIDEXG/2022/17, Project iTENTE, and \emph{e)} HORIZON-SESAR-2023-DES-ER-02 project ANTENNAE (101167288).}

} 

\maketitle
%\IEEEpeerreviewmaketitle

\begin{abstract}

We propose a data-driven approach for large-scale cellular network optimization, using a production cellular network in London as a case study and employing Sionna ray tracing for site-specific channel propagation modeling. We optimize base station antenna tilts and half-power beamwidths, resulting in more than double the 10\%-worst user rates compared to a 3GPP baseline. In scenarios involving aerial users, we identify configurations that increase their median rates fivefold without compromising ground user performance. We further demonstrate the efficacy of model generalization through transfer learning, leveraging available data from a scenario source to predict the optimal solution for a scenario target within a similar number of iterations, without requiring a new initial dataset, and with a negligible performance loss.

% Previous version
% We address the complex problem of designing cellular networks for 3D aerial corridors through a data-driven method. Utilizing high-dimensional Bayesian optimization (HD-BO), we simultaneously optimize cell antenna tilts and half-power beamwidths (HPBWs). Our findings reveal that this approach yields more than a fivefold increase in median UAV data rates along the 3D aerial corridors, while also improving ground user performance by 50\%, compared to a baseline configuration of an operational cellular network where all antennas are down-tilted. Additionally, we investigate the generalization capabilities of HD-BO via transfer learning, where data from an initially observed scenario (source) is used to predict optimal solutions for a new scenario (target). Remarkably, even without prior knowledge of the target scenario and relying solely on the source scenario data, the performance only reduces by a marginal 1\%, with convergence occurring within a comparable number of iterations.
\end{abstract}
\section{Introduction}
\label{sec:Intro}

The large-scale optimization of cellular networks remains a significant challenge due to the complex interdependencies among settings across multiple cells. Coverage and capacity are heavily influenced by the configuration of base station (BS) antennas, with adjustments in parameters such as the tilt angle and half-power beamwidth (HPBW) being critical for optimizing signal strength and minimizing interference. This process, known as cell shaping, becomes increasingly difficult in large-scale networks, as interactions between cells lead to a non-convex and NP-hard optimization problem \cite{tekgul2023joint}.

The conflicting goals of maximizing both coverage and capacity further complicate cellular network design. Coverage optimization involves directing energy toward cell edges. Capacity optimization prioritizes the signal-to-interference-plus-noise ratio (SINR) for users closer to the cell center. These challenges are exacerbated in scenarios involving nonhomogeneous user distributions, where uniform down-tilt configurations become ineffective \cite{GerGarAza2022,geraci2022integrating, benzaghta2022uav}.

Traditional methods for optimizing cellular networks, such as those employed in 3GPP frameworks, rely heavily on stochastic simulations and are typically limited to regular hexagonal deployments \cite{3GPP36814}. In real networks, site-specific radio frequency planning tools rely on time-consuming trial-and-error methods. 
A model-based framework for cellular network optimization, based on quantization theory and recently proposed in \cite{karimi2024optimizing}, enables the fine-tuning of antenna parameters for each BS within a given deployment to achieve optimal coverage, capacity, or any trade-off thereof. 

In this paper, we propose an alternative approach that leverages available data to maximize real-world key performance indicators (KPIs), which are often mathematically intractable. Our main contributions can be summarized as follows:

\subsubsection*{Data-driven optimization}

We employ high-dimensional Bayesian Optimization (HD-BO) to address a practical large-scale mobile network optimization problem, using a production cellular network in London as a case study. We employ a 3D representation of the area and use Sionna RT \cite{hoydis2023sionna} to model site-specific channel propagation, considering the actual cell locations and configuration. To evaluate the effectiveness of our data-driven approach, we jointly optimize antenna tilts and HPBWs and identify configurations that achieve more than double the $10$\%-worst rates with respect to a 3GPP baseline.

\subsubsection*{Aerial connectivity}

In a second case study, we maximize user rates on the ground as well as along 3D aerial corridors for uncrewed aerial vehicle (UAV) users. Ensuring reliable air-to-ground connectivity demands configurations that cannot be easily determined through heuristic methods \cite{maeng2023base,bernabe2024massive}. Our approach identifies configurations that improve UAV median rates by fivefold, without degrading ground performance.

\subsubsection*{Transfer learning}

In alignment with the 3GPP vision on data-driven model generalization \cite{3GPP38.843}, 
%we explore the capabilities of our approach in the context of \emph{transfer learning}. 
we use transfer learning to leverage knowledge from a previously optimized scenario (\emph{scenario source}), to predict the optimal solution for a new scenario (\emph{scenario target}). We demonstrate that, when the aerial corridor height changes from 140\,m--160\,m to 40\,m--60\,m, convergence occurs within a similar number of iterations, without the need for a new initial dataset, and with only a marginal 1\% performance decline.

\section{System Model}
\label{sec:Sys_Model}

In this section, we present the network deployment, channel model, and performance metrics utilized in our study.

\subsubsection*{Cellular network topology} 

\begin{figure}
\centering
\includegraphics[width=\figwidth, height=0.35\textheight, keepaspectratio]{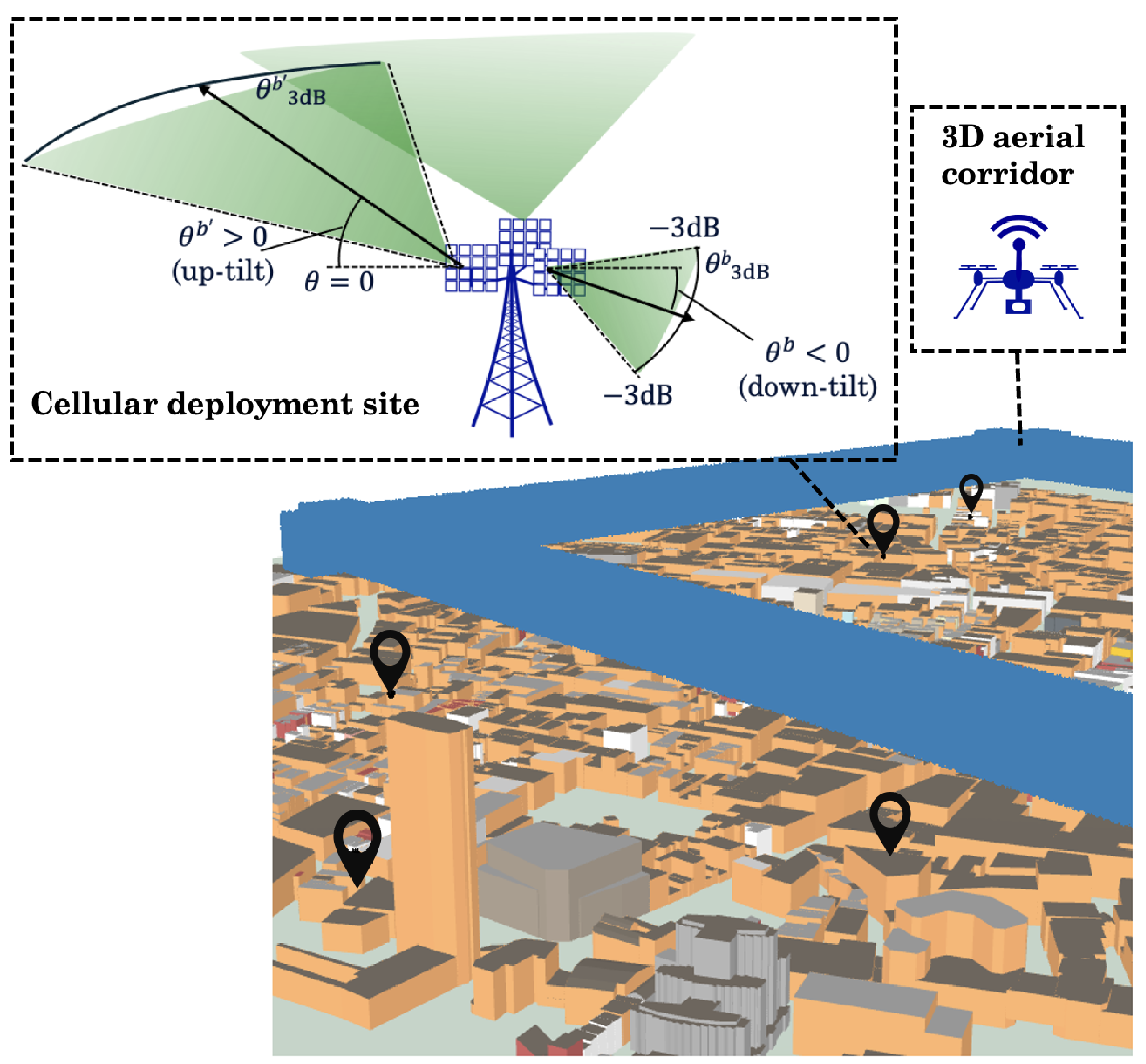}
\caption{A section of the area considered, with cell deployment sites indicated by black markers and 3D aerial corridors shown in blue.}
\label{fig:RD_illustration}
\end{figure}

We consider a site-specific scenario corresponding to a real-world production radio network owned by a leading commercial mobile operator in the UK. The network segment under consideration includes 16 deployment sites, with heights ranging from 22 to 56 meters. 
Each site consist of three sector antennas, resulting in a total of 48 cells across the network. The geographical area selected for our study covers approximately  $1400$\,m by $1275$\,m and is situated in London, between latitudes $[51.5087, 51.5215]$ and longitudes $[-0.1483, -0.1296]$. Fig.~\ref{fig:RD_illustration} illustrates a 3D model of the selected area, highlighting several cell site locations.

%\green{Within this area, the GUEs are randomly positioned outdoors (i.e., not within buildings) at a height of 1.5\,m, with an average density of 10 GUEs per cell \cite{3GPP36777}. Additionally, we consider four 3D aerial corridors within the area, each measuring 900\,m in length, 40\,m in width, and positioned at heights between 140\,m and 160\,m. The ratio of UAVs to GUEs is set at 50\%, in accordance with 3GPP Case~5 as detailed in \cite{3GPP36777}.} Figure \ref{fig:RD_illustration} illustrates a 3D model of the selected area in London, highlighting several cell site locations \green{and providing a representation of the 3D aerial corridors}.

\subsubsection*{Antenna model}

We characterize the antenna configuration of each BS $b$ by four main parameters: tilt $\theta^{b}$, bearing $\phi^{b}$, vertical HPBW $\theta^{b}_{\text{3dB}}$, and horizontal HPBW $ \phi^{b}_{\text{3dB}}$. The tilt is defined as the angle between the antenna boresight and the horizon and can be electrically adjusted. The bearing indicates the horizontal orientation of each sector. The vertical HPBW is the angular range over which the antenna gain is above half of the maximum gain in the vertical plane, while the horizontal HPBW is the corresponding range in the horizontal plane. In this study, the bearings $\phi^{b}$ and the horizontal HPBWs $\phi^{b}_{\text{3dB}}$ are assumed fixed: the former as per the real cellular deployment, the latter set to $\phi^{b}_{\text{3dB}}=65^{\circ}$ for all BSs. 
%
%\footnote{\hl{Are the bearings from the dataset? Or are they arbitrarily set to something like \{0, 120, 240\}?} \blue{[Md]: From the dataset.}}
%
The tilts $\theta^{b}$ and HPBWs $\theta^{b}_{\text{3dB}}$ are the object of optimization. 
The normalized antenna gain for a specific pair of azimuth and elevation, $\phi$ and $\theta$, between a BS $b$ and a UE $k$, is \cite{3GPP36814}, and where the maximum antenna gain depends on the HPBW (e.g., $\theta^{b}_{\text{3dB}}=10^{\circ}$ and $\phi^{b}_{\text{3dB}}=65^{\circ}$ yield a maximum gain of 14\,dBi \cite{3GPP36814}).

\vspace{-1em}
\begin{equation}
A(\phi, \theta)_{\textrm{dB}} = - \min \left\{ - \left[ A_{\text{H}}(\phi) + A_{\text{V}}(\theta) \right], 25 \right\} \, ,
\label{eqn: AG}
\end{equation}
where
\begin{equation}
A_{\text{H}}(\phi)_{\textrm{dB}} = - \min \left\{ 12 \left[ \left(\phi - \phi^{b}\right)/\phi_{\text{3dB}} \right]^2, 25 \right\} \, ,
\end{equation}
\begin{equation}
A_{\text{V}}(\theta)_{\textrm{dB}} = - \min \left\{ 12 \left[ \left(\theta - \theta^{b}\right)/\theta^{b}_{\text{3dB}} \right]^2, 20 \right\} \,.
\end{equation}

\subsubsection*{Site-specific propagation channel} 
A 3D representation of the selected area is constructed using OpenStreetMap, incorporating terrain and building information. The BSs are positioned and configured according to the actual cellular network topology. The omnidirectional large-scale channel gain (excluding the antenna gain) between BS $b$ and UE $k$ is calculated using Sionna RT \cite{hoydis2023sionna}, a widely-used 3D ray-tracing tool for analyzing site-specific radio wave propagation. Simulations are conducted at a carrier frequency of $2$\,GHz. The material \texttt{itu_concrete} is used to model the permittivity and conductivity of all buildings. The maximum number of reflections and diffractions are set to 5 and 1, respectively. The total large-scale channel gain $G_{b,k}$ is then obtained from the omnidirectional ray tracing channel gain by adding the antenna gain as per \eqref{eqn: AG}.

%%%%%%%%%%%%%%%%%%%%%%%%%%%%%%%%%%%%%%%%%%%%%%%%%%%

\subsubsection*{User distribution}

We denote by $\ncalU$ the set of users under consideration. User locations are drawn from a desired 3D  distribution, e.g., with nonhomogeneous density to model performance prioritization across regions (see Section~III-C).
%
%Davd: Indicate that more details will be given in Section III.C. 

%%%%%%%%%%%%%%%%%%%%%%%%%%%%%%%%%%%%%%%%%%%%%%%%%%%

\subsubsection*{SINR and achievable rates}

Through system-level simulations, we compute the downlink SINR in dB experienced by UE $k$ from its serving BS $b_k$ on a given time-frequency physical resource block (PRB), given by
\begin{equation}
  \sinr_{\textrm{dB},k} = 10\,\log_{10} \left( \,\frac{p_{b_k} \cdot G_{b_k,k}}{
  \sum\limits_{b\in\calB\backslash b_k}{p_{b} \cdot G_{b,k}  \,+\, \sigma_{\textrm{T}}^2}}\right),
  \label{SINR_DL_TN}
\end{equation}
where $\sigma_{\textrm{T}}^2$ denotes the thermal noise power and $p_{b}$ denotes the transmit power of BS $b$ on a PRB. 
The rate $\ncalR_k$ achievable by user $k$ served by BS $b_k$ can be related to its SINR as
\begin{equation}
    \ncalR_k = \eta_k B_k \, \mathbb{E} \left[ \log_2 (1 + \sinr_{k}) \right],
    \label{rates}
\end{equation}
where $B_k$ is the bandwidth allocated to user $k$, $\eta_k$ the fraction of time user $k$ is scheduled by the serving BS $b_k$, and the expectation is taken with respect to the small-scale fading.

\subsubsection*{Problem formulation}

Our goal is to maximize the rates in (\ref{rates}) for all UEs $k$ in the set $\ncalU$. 
We define the objective function:
\begin{equation}
    f\left(\bm{\theta}, \bm{\theta}_{\text{3dB}}\right) = \sum\limits_{k\in\ncalU}{\log  \ncalR_k}
    \label{eqn:Opt_problem_joint}
\end{equation}
\noindent to be maximized with respect to $\bm{\theta}$ and $\bm{\theta}_{\text{3dB}}$. The vector $\bm{\theta}$ contains the antenna tilt $\theta^{b} \in [-90^{\circ},90^{\circ}]$ of all BSs $b$ $\in$ $\calB$, with negative and positive angles denoting down-tilts and up-tilts, respectively. Similarly, the vector $\bm{\theta}_{\text{3dB}}$ contains the HPBW $\theta^{b}_{\text{3dB}} \in [0^{\circ},360^{\circ}]$ for each BS. 
While our data-driven approach can be employed to maximize any desired KPI---e.g., any function of the received signal strength, SINR, and rate---we selected the sum-log-rate in (\ref{eqn:Opt_problem_joint}) due to its widespread use in cellular systems to achieve fairness among users \cite{kelly1998rate}.

Maximizing \eqref{eqn:Opt_problem_joint} is a complex task since it requires tailoring the cellular deployment to the site-specific signal propagation patterns, while considering inter-cell interference and load balancing. Moreover, the users in $\ncalU$ may be distributed on an arbitrary 3D region with a nonhomogeneous density, thereby requiring performance prioritization across regions. 

\section{Data-driven Optimization}
\label{sec:HDBO}

%We tackle the challenge of optimizing large-scale cellular networks through a data-driven Bayesian Optimization (BO) approach.
We tackle the maximization of \eqref{eqn:Opt_problem_joint} through a data-driven Bayesian Optimization (BO) approach. 
Although BO \cite{shahriari2015taking} has previously proven useful in addressing coverage/capacity tradeoffs, optimal radio resource allocation, and mobility management \cite{dreifuerst2021optimizing, zhang2023bayesian, maggi2021bayesian, tambovskiy2022cell, tekgul2023joint, de2023towards,benzaghta2023designing}, it faces limitations due to the number of decision variables it can efficiently handle---typically around twenty or fewer in continuous domains \cite{frazier2018tutorial}. 
%---which limits the size of a cellular network and the number of parameters that can be optimized. 
In this paper, we take the first step towards employing high-dimensional BO (HD-BO) for optimizing large-scale cellular networks, thus overcoming the limitations of traditional BO.%
% Other reference: maggi2023energy
%
%
\footnote{While reinforcement learning (RL) is also promising, % \cite{BalAnd2019,VanIakHak2021}, 
it tends to require a very large amount of data and to converge slowly. RL also employs random exploration, which in a real production network can lead to test highly suboptimal antenna configurations that degrade the system performance \cite{tekgul2023joint}.}
% Other refs on RL: \cite{DanShaKle2017,BouFarFor2021}
%

\subsection{Introduction to Bayesian Optimization}

BO is a suitable framework for black-box optimization, where the objective function $f(\cdot)$ is non-convex, non-linear, stochastic, and/or computationally expensive to evaluate. BO operates by iteratively building a probabilistic \textit{surrogate model} of $f(\cdot)$ from previous evaluations at selected query points \cite{shahriari2015taking}. 

We define a query point $\textbf{x} = [\bm{\theta}, \bm{\theta}_{\text{3dB}}]$ as a configuration of the antenna tilts $\theta^{b}$ and vertical HPBW $\theta^{b}_{\text{3dB}}$ of each BS $b$ $\in$ $\calB$, and obtain the corresponding value of $f(\textbf{x})$ from (\ref{eqn:Opt_problem_joint}). Since $f(\cdot)$ is a mathematically intractable function---capturing the site-specific propagation channel and the inherent stochasticity of the UE locations---we evaluate $f(\cdot)$ through system-level simulations where the 3D locations of the UEs in $\ncalU$ are sampled at random according to a desired distribution and the channel is generated via Sionna RT. Each evaluation at a point $\textbf{x}$ yields a noisy sample $\tilde{f}(\textbf{x})$.
%
\begin{comment}
\footnote{
\hl{I think there has to be some randomness in $\ncalU$ in (6). \\ Option 1. We could define the rates on a continuous region, and then sample points of this region. But then the units would not match, because the rate over a region is bits/sqm not bits/UE. \\ Option 2. We could still define the set $\ncalU$ as discrete but random, following a certain density distribution of users (which we need to specify). Then I am not sure whether $f()$ should be also random or it should rather be the expectation over $\ncalU$ and thus deterministic.}\blue{[Md]: What we actually did is option 1, then we mapped those bits/sqm to bits/UE considering 3GPP BS deployment density per sqm and user deployment case 5. This is because we are using the coverage map in this study, where I have a grid per every meter. This is actually equivalent to option 2, it will be the same if we deploy the users randomly, measure the performance and take the expectation over several drops. I think in this case $f()$ is not random anymore, so I should remove the noise symbol from the paper. Before when using 3GPP channel model if I deploy a user in a specific location again, it will have a variance in the RSRP because of shadowing and LOS condition probabilistic models, that's why I was using the noisy $f()$. Now in Sionna, deploying the user again in the same location will not lead to a noisy sample, it is the same response.}}
\end{comment}
%
In practice, these samples could also be obtained through real-time measurements. 

For convenience, let us define $\textbf{X} = [\textbf{x}_1,\ldots,\textbf{x}_N]$ as a set of $N$ query points 
and $\tilde{\textbf{f}}(\textbf{X})=[\tilde{f}_1,\ldots,\tilde{f}_N]^\top$ as the set of corresponding evaluations, with $\tilde{f}_i = \tilde{f}(\textbf{x}_i)$, $i=1,\ldots,N$. We generate an initial dataset $\mathcal{D} = \{\textbf{x}_1,\ldots,\textbf{x}_{N_{\textrm{o}}},\tilde{f}_1,\ldots,\tilde{f}_{N_{\textrm{o}}}\}$ containing $N_{\textrm{o}}$ initial observations and constructed according to the model and objective function in Section~\ref{sec:Sys_Model}. From $\mathcal{D}$, we use a Gaussian process (GP) prior, $\widehat{f}(\cdot)$, to create a surrogate model (i.e., the posterior) that approximates the objective function, $\tilde{f}(\cdot)$ \cite{shahriari2015taking}. The GP model allows to predict the value of $\tilde{f}(\textbf{x})$ for a query point $\textbf{x}$ given the previous observations $\tilde{\textbf{f}}(\textbf{X})=\tilde{\textbf{f}}$ over which the model is constructed. Formally, the GP prior on $\tilde{f}(\textbf{x})$ prescribes that, 
for any set of inputs $\textbf{X}$,
the corresponding objectives $\tilde{\textbf{f}}$ are jointly distributed as
\begin{equation}
  p(\,\tilde{\textbf{f}}\,) = \mathcal{N}(\,\tilde{\textbf{f}} \,\,|\,\, \boldsymbol{\mu}(\textbf{X}),\mathbf{K}(\textbf{X})\,),
  \label{posterior}
\end{equation}
where $\boldsymbol{\mu}(\textbf{X}) = [\mu(\mathrm{\textbf{x}}_1),\ldots,\mu(\mathrm{\textbf{x}}_N)]^\top$ is the $N \times 1$ mean vector, 
and $\mathbf{K}(\textbf{X})$ is the $N \times N$ covariance matrix, 
whose entry $(i,j)$ is given by the covariance $k(\textbf{x}_{i},\textbf{x}_{j})$. 
%whose entry $(i,j)$ is given as $[\textbf{K}(\textbf{X})]_{i,j} = k(\textbf{x}_{i},\textbf{x}_{j})$ with $i,j \in \{1,\hdots,N\}$. 
For a point $\textbf{x}$, 
the mean $\mu(\textbf{x})$ provides a prior knowledge on $\tilde{f}(\textbf{x})$, 
while the kernel $\mathbf{K}(\textbf{X})$ indicates the uncertainty across pairs of values of \textbf{x}. 

Given a set of observed noisy samples $\tilde{\textbf{f}}$ at previously sampled points $\textbf{X}$, the posterior distribution of $\widehat{f}(\textbf{x})$ at point $\textbf{x}$ can be obtained as \cite{frazier2018tutorial} 
\begin{equation}
  p(\widehat{f}(\textbf{x}) = \widehat{f} \,\, | \,\, \textbf{X}, \textbf{$\tilde{\textbf{f}}$} \,) = \mathcal{N}(\widehat{f} \,\,|\,\, \mu(\textbf{x} \,\,|\,\, \textbf{X}, \tilde{\textbf{f}}),\sigma^2(\textbf{x} \,\,|\,\, \textbf{X}, \tilde{\textbf{f}})),
  \label{posterior_Noisy}
\end{equation}
with
\begin{equation}
  \mu(\textbf{x} \,|\, \textbf{X},\tilde{\textbf{f}}) = \mu(\textbf{x}) + \tilde{\textbf{k}}(\textbf{x})^\top (\tilde{\textbf{K}}(\textbf{X}))^{-1}(\tilde{\textbf{f}}-\boldsymbol{\mu}(\textbf{X})),
  \label{Mean_posterior_Noisy}
\end{equation}
\begin{equation}
  \sigma^2(\textbf{x} \,|\, \textbf{X},\tilde{\textbf{f}}) = k(\textbf{x},\textbf{x}) - \tilde{\textbf{k}}(\textbf{x})^\top (\tilde{\textbf{K}}(\textbf{X}))^{-1} \,\tilde{\textbf{k}}(\textbf{x}),
  \label{Kernel_posterior_Noisy}
\end{equation}
where 
$\tilde{\textbf{k}}(\textbf{x}) = [k(\textbf{x},\textbf{x}_{1}),\ldots,k(\textbf{x},\textbf{x}_{N})]^\top$ is the $N \times 1$ covariance vector and $\tilde{\textbf{K}}(\textbf{X}) = \textbf{K}(\textbf{X}) + \sigma^2 \textbf{I}_{\text{N}}$, with $\sigma^2$ denoting the observation noise represented by the variance of the Gaussian distribution, and $\textbf{I}_{\text{N}}$ denoting the $N \times N$ identity matrix.

An \textit{acquisition function} $\alpha(\cdot)$ (i.e., Thompson sampling) is then employed to score the response from the surrogate model (i.e., the posterior) and determine which point in the search space should be evaluated next. 

%The acquisition function balances exploration (searching for new and potentially better solutions) and exploitation (focusing on refining the current best solutions).

% This part was repeated:
% Although BO has proven useful in addressing coverage/capacity tradeoffs, optimal radio resource allocation, and mobility management \cite{benzaghta2023designing, dreifuerst2021optimizing, eller2024differentiable, zhang2023bayesian, maggi2021bayesian, tambovskiy2022cell, maggi2023energy, tekgul2023joint, de2023towards}, it faces significant limitations in terms of scalability, due to the number of decision variables it can efficiently handle. The computational complexity increases significantly with the dimensionality of the problem, making it less efficient for high-dimensional optimization tasks, particularly when the number of optimization parameters exceeds 20 variables \cite{eriksson2019scalable}. This constrains the size of the cellular network and the number of antenna parameters that can be optimized. 

Traditional BO faces limitations in terms of scalability, due to the limited number of decision variables it can efficiently handle. For the maximization of \eqref{eqn:Opt_problem_joint}, this constrains the number of antenna parameters that can be jointly optimized.

\subsection{High-dimensional Bayesian Optimization}

For BO methods to become more sample-efficient for a larger number of decision variables, it is essential to introduce a hierarchical significance for the dimensions. 
High-dimensional BO (HD-BO) leverages the fact that certain features may play a crucial role in capturing the behavior of $f(\cdot)$, while others may be of negligible importance.

We employ a state-of-the-art HD-BO method, Trust Region BO (TuRBO) \cite{eriksson2019scalable}, to tackle large-scale cellular network design.%
\footnote{We implemented and tested three HD-BO methods: Sparse Axis-Aligned Subspaces (SAASBO) \cite[Section~4]{eriksson2021high}, BO via Variable Selection (VSBO) \cite[Section~3]{shen2021computationally}, and Trust Region BO (TuRBO) \cite[Section~2]{eriksson2019scalable}. Due to space constraints, our discussion will focus on TuRBO, since it demonstrated superior performance and higher suitability for the problem under consideration.}
TuRBO transitions from global surrogate modeling to the management of multiple independent local models, with each model concentrating on a distinct region of the search space. TuRBO attains global optimization by simultaneously operating several local models and strategically allocating samples through an implicit multi-armed bandit approach. This enhances the acquisition strategy's effectiveness by focusing samples on promising local optimization endeavors.

TuRBO integrates trust region (TR) methods from stochastic optimization, which are gradient-free and utilize simple surrogate models within a defined TR, typically represented as a sphere or polytope centered around the best identified solution. However, these simple surrogate models may necessitate overly small trust regions for precise modeling. To overcome this limitation, TuRBO employs a GP surrogate model within the TR, maintaining essential global BO characteristics such as noise robustness and systematic uncertainty management.

In TuRBO, the TR is defined as a hyperrectangle centered at the current optimal solution, $f^{*}$. The side length of the TR is initialized as $L \gets L_{\text{init}}$. Subsequently, the side length for each dimension $L_i$ is adjusted based on its respective length scale $\lambda_i$ in the GP model. The side length for each dimension is then specified by:
\begin{equation}
L_i = {\lambda_i \, L}\cdot{\left(\sideset{}{_{j=1}^d}\prod \lambda_j \right)^{-1/d}}\!\!\!\!\!\!.
\end{equation}
where $d$ is the total number of dimensions (i.e, optimization parameters under consideration). This approach ensures that the TR adapts to the local characteristics of the search space, facilitating more effective optimization in high-dimensional settings.

During each local optimization run, an acquisition function selects a batch of  $q$ candidates at each iteration, ensuring they remain within the designated TR. If the TR's side length $L$ were large enough to cover the entire search space, this method would be equivalent to standard global BO. Thus, adjusting $L$ is crucial: the TR needs to be large enough to encompass good solutions but compact enough to ensure the local model's accuracy. The TR is dynamically resized based on optimization progress: it is doubled ($L \gets \min \{L_{\text{max}}, 2L\}$) after $\tau_{\text{succ}}$ consecutive successes, and halved ($L \gets L/2$) after $\tau_{\text{fail}}$ consecutive failures. Success and failure counters are reset after adjustments. If $L$ falls below $L_{\text{min}}$, that TR is discarded and a new one is initiated at $L_{\text{init}}$. The TR's side length is capped at $L_{\text{max}}$. 

In this study, we run TuRBO using an open-source repository \cite{eriksson2019scalable} with: $\tau_{\text{succ}} = 3$, $\tau_{\text{fail}} = 15$, $L_{\text{init}} = 0.8$, $L_{\text{min}} = 2^{-7}$, and $L_{\text{max}} = 1.6$. 
Thompson sampling is used as an acquisition function for selecting candidates both within and across TRs. TuRBO maintains $q=4$ candidates from the union of all trust regions.

\subsection{Case Studies}

To evaluate the effectiveness of our HD-BO approach, we jointly optimize antenna tilts and HPBWs to maximize the sum-log-rate in (\ref{eqn:Opt_problem_joint}). 
%The main system-level parameters are listed in Table~\ref{table:parameters}. 
We study two scenarios, as follows:

\subsubsection*{Case study \#1 --- Ground users only}

Within the area under consideration, a set $\ncalG$ of ground users (GUEs) are randomly positioned at uniform outdoors (i.e., not within buildings) at a height of 1.5\,m, with an average density of 10 GUEs per cell \cite{3GPP36777}. In this first case study, we set $\ncalU = \ncalG$.

\subsubsection*{Case study \#2 --- Ground users and UAV corridors}

Besides the set of GUEs $\ncalG$, we consider a set $\ncalA$ of uncrewed aerial vehicle (UAV) users in four 3D aerial corridors within the area, each measuring 900\,m in length, 40\,m in width, and positioned at heights between 140\,m and 160\,m (see Fig.~\ref{fig:RD_illustration}). 
In this second case study, we set $\ncalU = \ncalG \cup \ncalA$, to optimize both GUE and UAV performance.
%allowing to trade off GUE and UAV performance by choosing the cardinalities of $\ncalG$ and $\ncalA$ accordingly. 
We set the ratio of UAVs to GUEs at 50\%, in accordance with 3GPP Case~5 in \cite{3GPP36777}. 

%%%%%%%%%%%%%%%%%%%%%%%%%%%%%%%%%%%%%%%%%%%%%%%%%%%%%%%%
\begin{table}
\centering
\caption{System-level parameters for the case studies}
\label{table:parameters}
\def\arraystretch{1.2}
\begin{tabulary}{\columnwidth}{ |p{2.1cm} | p{5.85cm} | }
\hline
%	\textbf{Deployment} 			&  \\ \hline
  Cellular layout				& Production radio network, 16 deployment sites at $22$--$56$~m, three sectors per site, one BS per sector \\ \hline
  Frequency band 		&  $B_{k}\!=\!$ $10$ MHz at $2$~GHz \\ \hline
  	Thermal noise 				& $-174$\,dBm/Hz density \\ \hline 
	BS max power 			& $46$\,dBm over the whole bandwidth \cite{3GPP36814}\\ \hline  \hline

%	\textbf{Users} 			&  \\ \hline
  User association				& Based on received signal strength \\ \hline

	%User antenna 		& Omnidirectional, gain: 0~dBi \\ 	
    User receiver 		& Omnidirectional antenna \\ \hline
    GUE distribution 				& $10$ per sector on average, outdoor, at $1.5$\,m \\ \hline

    UAV distribution 				& Uniform in four aerial corridors at $140$--$160$\,m height, $70$ UAVs per corridor on average %(see also Fig.~\ref{fig:RD_illustration}) 
    \\ \hline
      
      UAVs/GUEs ratio 				& $50$\% as per 3GPP Case~5 \cite{3GPP36777} \\ \hline

\end{tabulary}
\end{table}

% As it will be shown, ensuring connectivity along UAV corridors demands configurations that cannot be easily achieved through heuristic methods \cite{bernabe2024massive, maeng2023base,KarGerJafICC2023}.
 
% Figure \ref{fig:RD_illustration} illustrates a 3D model of the selected area in London, highlighting several cell site locations and providing a representation of the 3D aerial corridors.

%%%%%%%%%%%%%%%%%%%%%%%%%%%%%%%%%

%%%%%%%%%%%%%%%%%%%%%%%%%%%%%%%%%%%%%%%%%%%%%%%%%%%

%%%%%%%%%%%%%%%%%%%%%%%%%%%%%%%%%%%%%%%%%%%%%%%%%%%

\subsubsection*{Data-driven optimal performance}

\begin{figure}
\centering
\includegraphics[width=\figwidth]{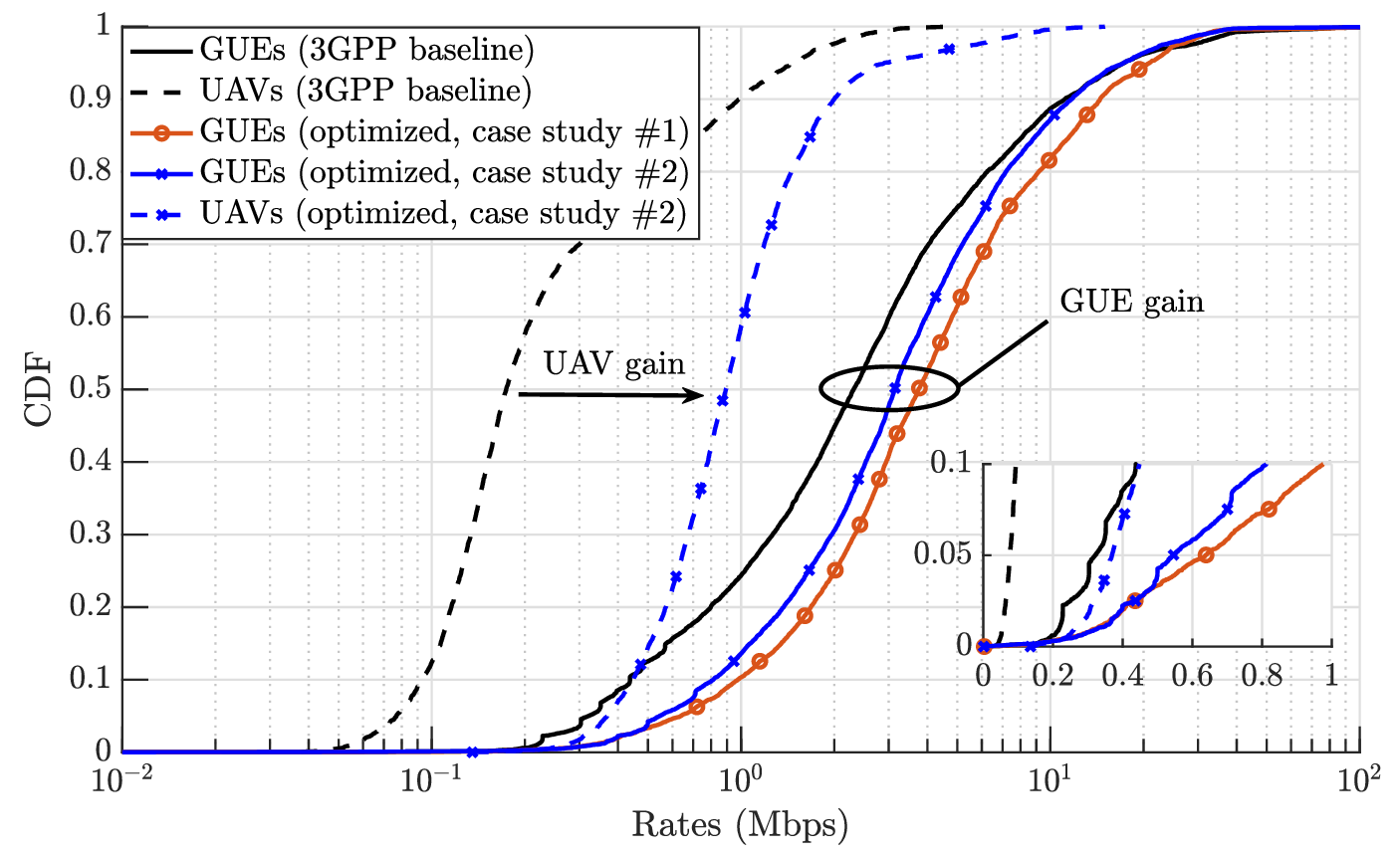}
\caption{Rates achieved by GUEs and UAVs when the network is optimized for GUEs only (case study \#1), GUEs and UAV corridors (case study \#2), and with a 3GPP baseline configuration.}
\label{fig:Rates_CDFs_tilts_vHPBW}
\end{figure}

Fig.~\ref{fig:Rates_CDFs_tilts_vHPBW} presents the cumulative distribution function (CDF) of rates for GUEs (solid) and UAVs (dashed).
Black lines represent the performance under the baseline 3GPP configuration, with uniform tilts $\theta^{b} = -12^{\circ}$ and vertical HPBW $\theta^{b}_{\text{3dB}} = 10^{\circ}$ $\forall b$ \cite{3GPP36814}.
The red line shows the GUE performance after data-driven optimization for GUEs only (case study \#1). The blue lines show the performance when optimizing for both GUEs and UAVs (case study \#2). The following observations can be made:
\begin{itemize}[leftmargin=*]
\item
For case study \#1, data-driven optimization of tilt and HPBW results in a 120\% and 63\% improvement in the 10\%-tile and median GUE rates, respectively, compared to the 3GPP configuration (solid red vs. solid black).
\item 
For case study \#2, under the baseline configuration, $12$\% of UAVs have data rates below $100$\,kbps, the minimum requirement set by 3GPP for command and control links \cite{3GPP36777} (dashed black).%
\footnote{While the Shannon rates in (\ref{rates}) are always non-zero, $80$\% of the UAVs have SINRs below $-5$\,dB: a proxy for outage \cite{geraci2018understanding}. The optimal tilt and HPBW configuration reduces this outage from $80$\% to zero.}
Data-driven optimization increases UAV rates nearly fivefold in both the $10$\%-tile and median (dashed blue vs. dashed black).
\item
For case study \#2, the improvement in UAV performance does not lead to a significant GUE performance degradation. 
The $10$\%-tile GUE rates are only $17$\% lower than those optimized for GUEs only (solid blue vs. solid red) and are still $82\%$ higher than the baseline 3GPP configuration (solid blue vs. solid black). 
This demonstrates the capability of the data-driven approach to identify optimal trade-offs across the 3D user region, encompassing ground and aerial users.
% On the other hand, even when the UAVs are considered as in case study \#2, the GUE rates show a 35\% increase in median compared to the baseline scenario.

\end{itemize}

%%%%%%%%%%%%%%%%%%%%%%%%%%%%%%%%%

\subsubsection*{Data-driven optimal configuration}

\begin{figure}
\centering
\includegraphics[width=\figwidth]{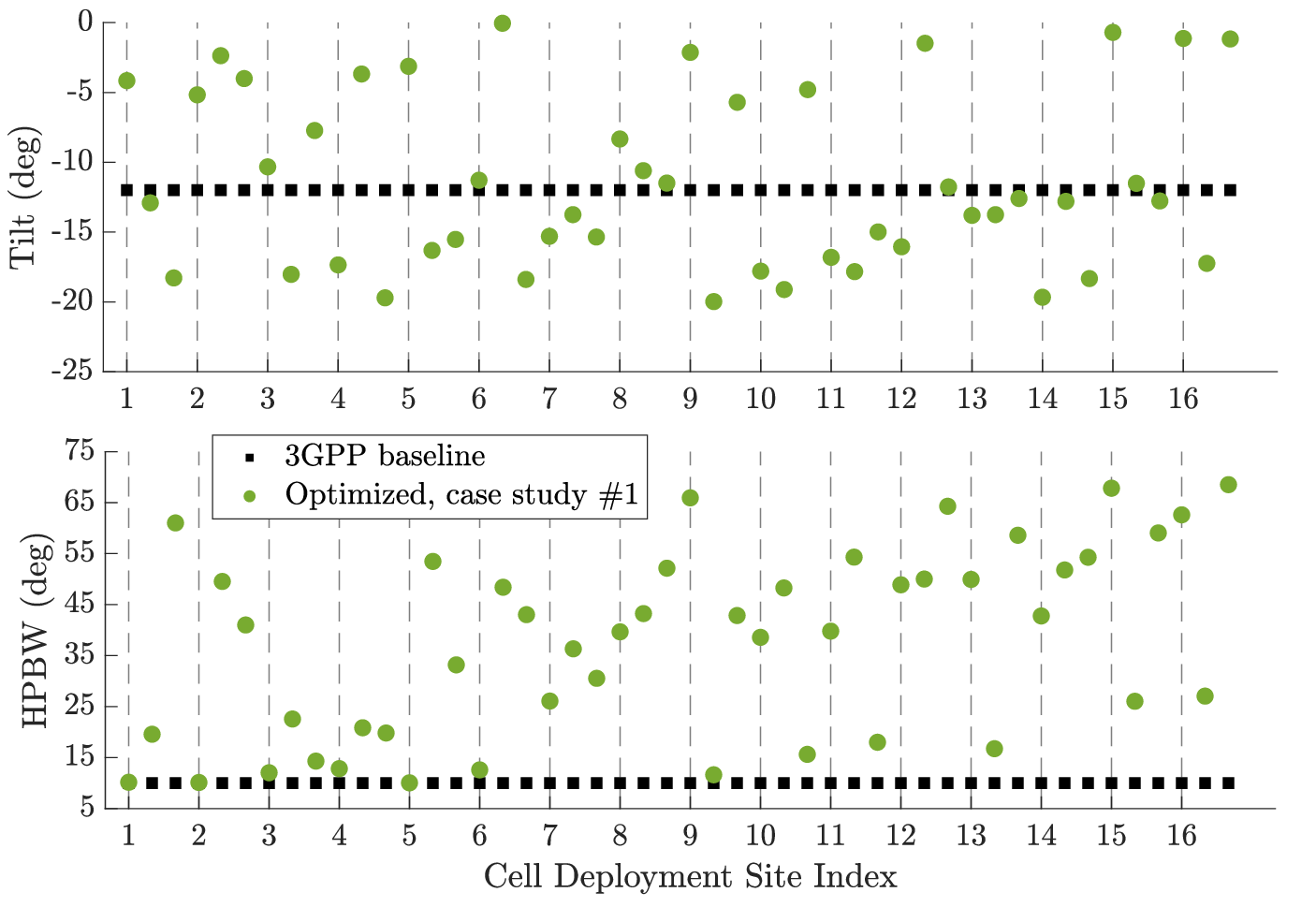}
\caption{Optimal tilts and HPBW for case study \#1, with GUEs only.}
\label{fig:Opt_Tilts_vHPBW_config_TuRBO_GUEs}
\end{figure}

\begin{figure}
\centering
\includegraphics[width=\figwidth]{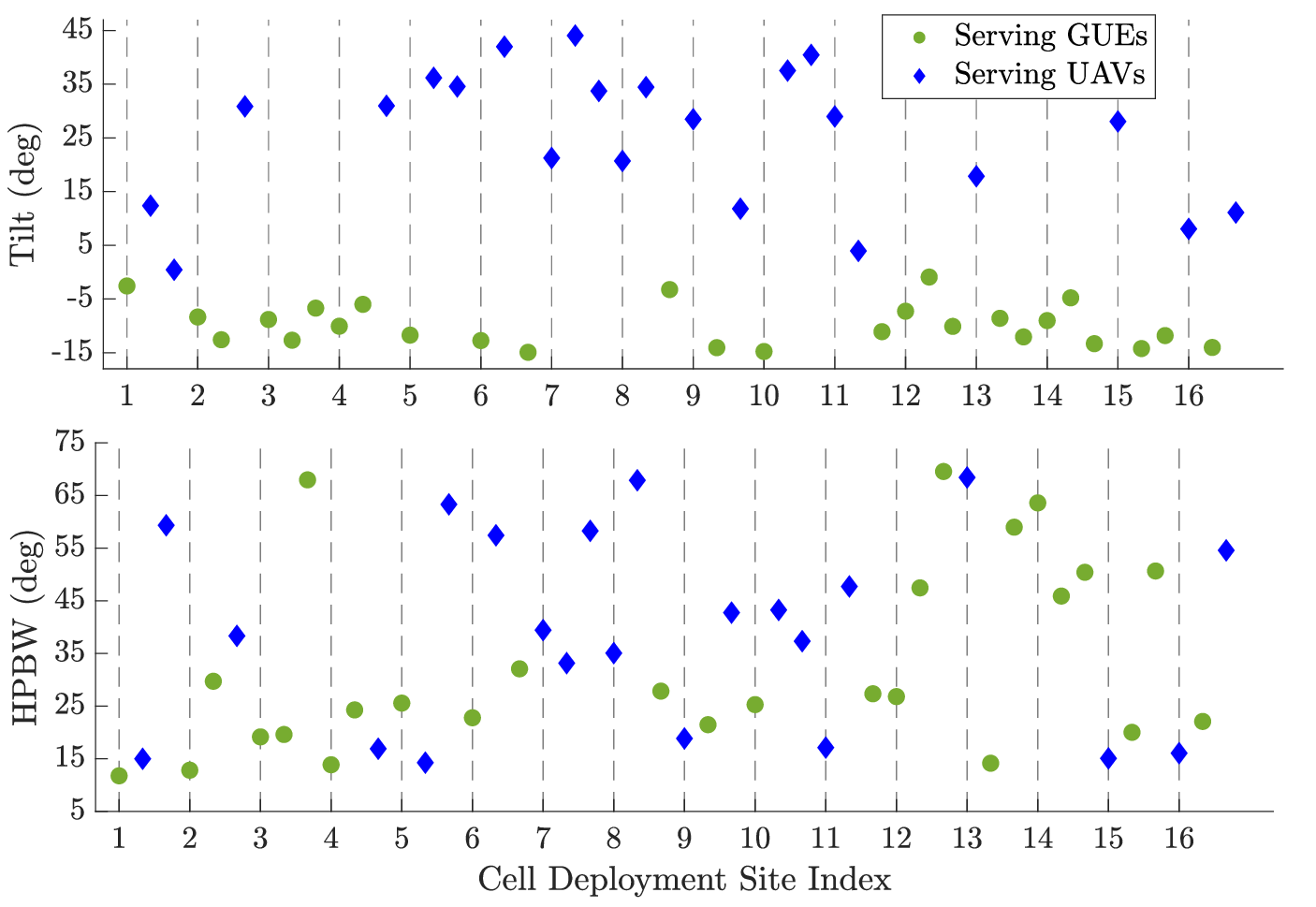}
\caption{Optimal tilts and HPBW for case study \#2. Green circles and blue diamonds denote BSs serving GUEs and UAVs, respectively.}
\label{fig:Opt_Tilts_vHPBW_config_TuRBO}
\end{figure}

Fig.~\ref{fig:Opt_Tilts_vHPBW_config_TuRBO_GUEs} and Fig.~\ref{fig:Opt_Tilts_vHPBW_config_TuRBO} show the optimal configurations of antenna tilts $\bm{\theta}$ and HPBWs $\bm{\theta}_{\text{3dB}}$ for the two case studies. Each index denotes the deployment site with three sectors. We note the following:
\begin{itemize}[leftmargin=*]
\item 
In case study \#1, as shown in Fig.~\ref{fig:Opt_Tilts_vHPBW_config_TuRBO_GUEs}, the optimal data-driven configuration (green dots) deviates significantly from the uniform 3GPP baseline (black squares), with BSs exhibiting varied tilts and often wider HPBWs than the baseline $10$º. This setup, tailored to the irregular urban deployment, leads to the performance gains highlighted in Fig.~\ref{fig:Rates_CDFs_tilts_vHPBW}.
\item 
In case study \#2, Fig.~\ref{fig:Opt_Tilts_vHPBW_config_TuRBO} uses markers to differentiate between cells serving GUEs (green circles) and UAVs (blue diamonds). Unlike case study \#1, where all BSs are down-tilted, optimizing for both GUEs and UAVs results in 22 BSs being up-tilted, with the remaining BSs down-tilted. 
While 18 BSs in Fig.~\ref{fig:Opt_Tilts_vHPBW_config_TuRBO_GUEs} have down-tilted angles below $-15$º, all down-tilted BSs in Fig.~\ref{fig:Opt_Tilts_vHPBW_config_TuRBO} maintain angles above $-15$º, effectively filling the coverage gaps left by the up-tilted BSs.
% Also in case study \#2 only six BSs have HPBWs narrower than $15$º.
%
%\footnote{\hl{In Fig. 4, did you force the tilts to be higher than -15º? Or is that the solution? Did you use the same range between Figs. 3 and 4?} \blue{[Md]: I only forced it in the upper bound to be 0deg. So for case1: [0,-20] and case2: [45,-20]}}
%
\end{itemize}

\noindent Overall, the non-trivial optimal configurations of tilts and HPBWs, particularly in case study \#2, reflect the complexity of achieving the best performance trade-offs.

\section{Transfer Learning}
\label{sec:TL}

 %Therefore, it is crucial to understand the conditions under which the model generalizes effectively \cite{lin2023overview}. 
%A generalist model, as defined, is one that can be applied to different scenarios, configurations, or sites \cite{lin2023overview}.
The commercial adoption of a machine learning model requires consistent performance across various scenarios \cite{lin2023overview}.
%, configurations, and site-specific conditions  
In this section, we explore the generalization capabilities of the HD-BO framework across 
%various scenarios, specifically 
different UE distributions, within the context of \emph{transfer learning}.

\subsubsection*{Scenario source vs. scenario target}

Transfer learning in optimization utilizes knowledge or data from a previously solved problem (\emph{source}) to expedite the solution of a new but related problem (\emph{target}). {This method proves especially advantageous when generating the initial dataset $\mathcal{D}$ required for the BO posterior is costly or time-consuming, e.g., because it requires measurements. Let $\mathcal{D}_{\text{sr}}$ and $\mathcal{D}_{\text{tg}}$ denote initial datasets obtained for scenarios source and target, respectively. We conduct three evaluations, varying the percentage of the initial dataset $\mathcal{D}$ that is based on scenario target, as follows:
\begin{itemize}[leftmargin=*]
\item 
100\% ($\mathcal{D} = \mathcal{D}_{\text{tg}}$, prior knowledge based on scenario target).
\item 
50\% (half of $\mathcal{D}$ is drawn from 
$\mathcal{D}_{\text{sr}}$, half is from $\mathcal{D}_{\text{tg}}$).
\item 
0\% ($\mathcal{D} = \mathcal{D}_{\text{sr}}$, prior knowledge based on scenario source).
\end{itemize}
}

\begin{comment}
Three evaluations were conducted, varying the initial dataset's dependency on the scenario target: 
\begin{itemize}[leftmargin=*]
\item 
$100$\% of the initial dataset is taken from the scenario target (no transfer learning is performed).
\item 
$50$\% of the initial dataset is from the scenario target and $50$\% from the scenario source (partial transfer learning).
\item 
$0$\%, i.e., the initial dataset is taken entirely from the scenario source (full transfer learning).
\end{itemize}
\end{comment}

We apply \emph{scenario-specific} transfer learning to case study \#2, where our objective is to utilize data collected from applying data-driven optimization to a particular UAV corridors height to a new scenario where the height has changed. The scenario source is based on the previously described case study \#2, consisting of GUEs and UAVs along 3D aerial corridors at an altitude between $140$\,m--$160$\,m. The scenario target changes the aerial corridors height to $40$\,m--$60$\,m.

\subsubsection*{Convergence of transfer learning}

\begin{figure}
\centering
\includegraphics[width=\figwidth]{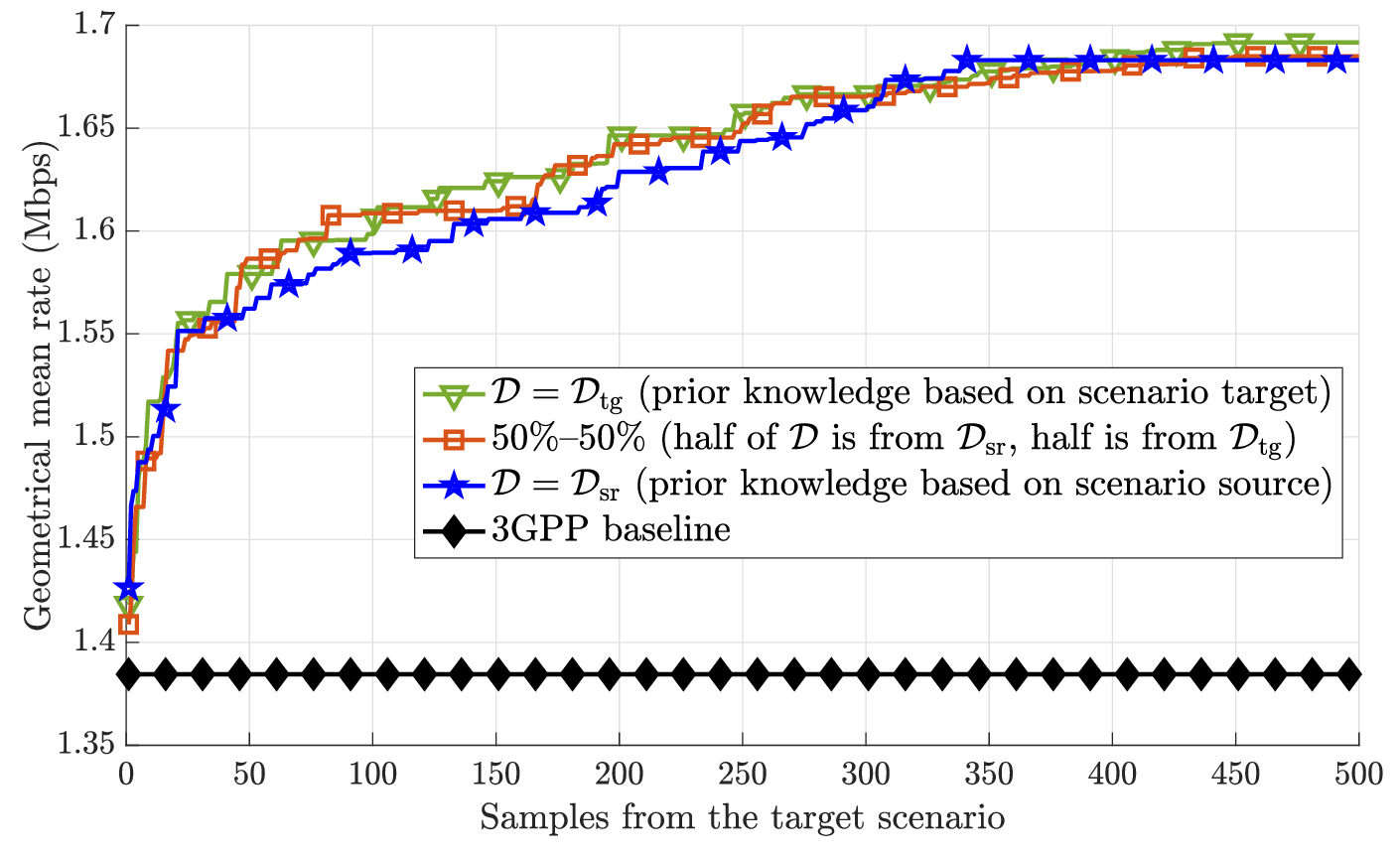}
\caption{Convergence of transfer learning applied on case study \#2. The initial dataset $\mathcal{D}$ contains $N_{\textrm{o}}=200$ observations.}
\label{fig:LT_conv_50m_150m}
\end{figure}

Fig.~\ref{fig:LT_conv_50m_150m} illustrates the convergence of transfer learning using HD-BO, showing the best observed objective at each iteration $n$. 
To show a quantity of practical interest, we plot the geometrical mean rate across all UEs in $\ncalU$, closely related to the objective function $f(\cdot)$ in (\ref{eqn:Opt_problem_joint}) as follows:
\begin{equation}
\overline{\mathcal{R}} = \left( \sideset{}{_{k\in\ncalU}}\prod \ncalR_k \right)^{\frac{1}{|\ncalU|}} = e^{{f(\cdot)}/{|\ncalU|}},
\end{equation}
where $|\ncalU|$ denotes the cardinality of the UE set. 
The initial dataset $\mathcal{D}$ contains $N_{\textrm{o}}=200$ observations drawn from $\mathcal{D}_{\text{sr}}$ (blue), from $\mathcal{D}_{\text{tg}}$ (green), or half each (red). Successive samples are collected on the target scenario (x-axis). 
Fig.~\ref{fig:LT_conv_50m_150m} shows that with a 50\%/50\% reliance on $\mathcal{D}_{\text{tg}}$/$\mathcal{D}_{\text{sr}}$, convergence occurs in a comparable number of iterations to that observed with 100\% reliance on $\mathcal{D}_{\text{tg}}$ (i.e., without transfer learning). This shows that resources can be conserved when generating the initial dataset $\mathcal{D}$, highlighting the HD-BO posterior's capability to generalize after completing an optimization run for a related task. Even without prior knowledge of the target ($\mathcal{D} = \mathcal{D}_{\text{sr}}$), performance declines by just $1$\%.

\subsubsection*{Performance of transfer learning}

\begin{figure}
\centering
\includegraphics[width=\figwidth]{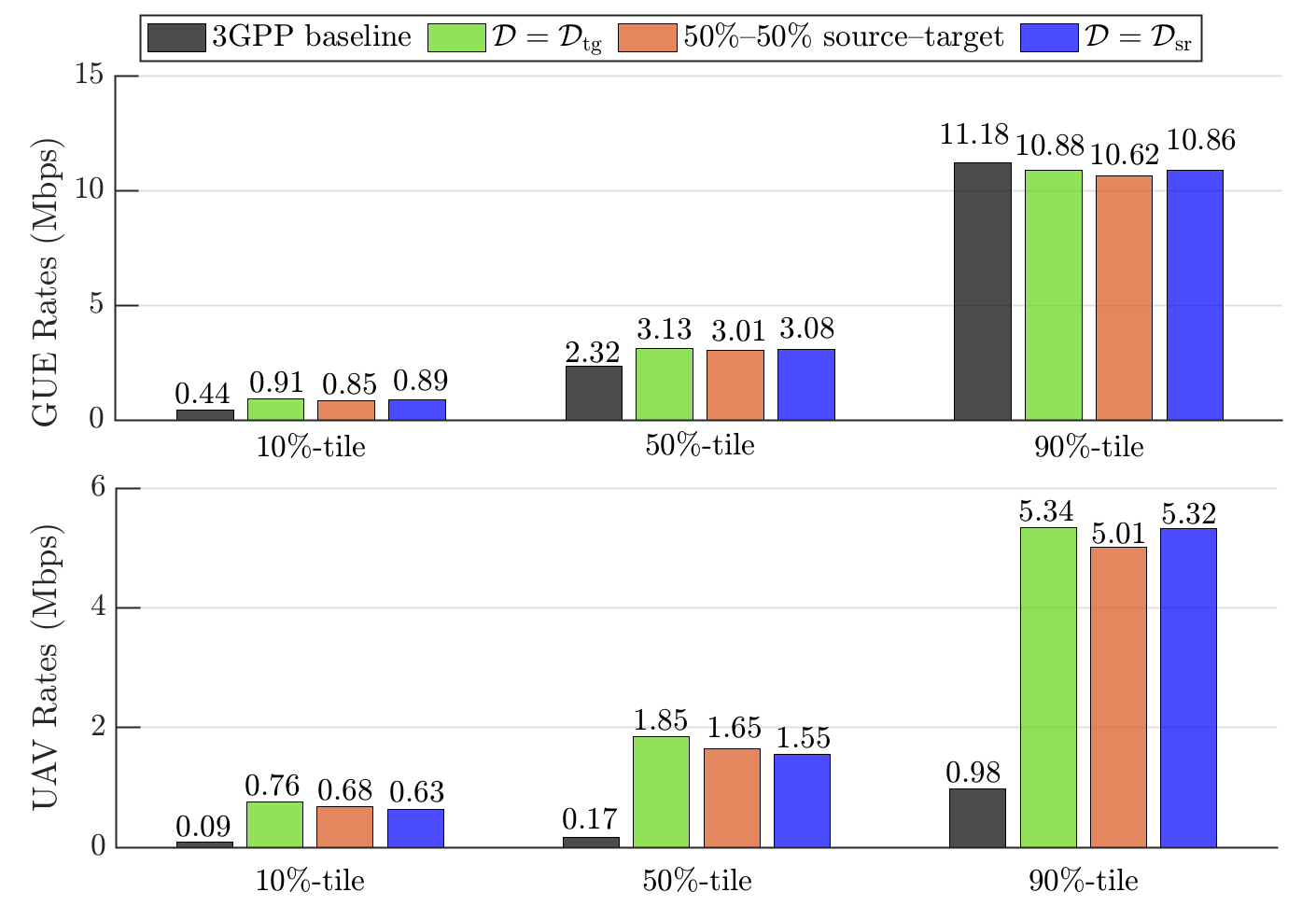}
\caption{Performance of transfer learning applied on case study \#2.}
\label{fig:LT_compare_50m_150m_bar}
\end{figure}

Fig.~\ref{fig:LT_compare_50m_150m_bar} shows the effectiveness of transfer learning in terms of achievable rates. 
Similarly to case study \#2, this figure shows that data-driven optimization of tilts and HPBWs significantly improves the UAV rates when these are distributed along aerial corridors of between $40$\,m and $60$\,m in height, with an order of magnitude gain in median UAV rates and $35\%$ gains in median GUE rates (green vs. gray). 
Importantly, under full transfer learning, the UAVs median rate is only reduced by $16$\% compared to data-driven optimization without transfer learning, while the GUE median rate is nearly preserved (blue vs. green).

% Previous longer version
% Fig.~\ref{fig:LT_compare_50m_150m_bar} shows the achieved rates for both GUEs and UAVs. The optimization through TuRBO significantly improves the UAV rates along aerial corridors height between 40\,m and 60\,m,  with 8$\times$, 10$\times$, and  5$\times$ gains for the 10\%-tile, 50\%-tile, and 90\%-tile, respectively, without causing a severe GUE performance degradation. It is also observed that with full transfer learning, the UAVs rate are only reduced by 16\% for the 10\%-tile, and 50\%-tile, while maintaining the same rates for the 90\%-tile, compared to the benchmark, in which no transfer learning is applied. This demonstrates the effectiveness of transfer learning in accommodating UAVs in new aerial corridor deployments while preserving performance for GUEs.

%%%%%%%%%%%%%%%%%%%%%%%%%%%
%%%%%%%%%%%%%%%%%%%%%%%%%%%

%We conclude this section with two remarks:

\begin{comment}
\subsubsection*{Applicability of transfer learning}

As one may expect, transfer learning does not always work. We also studied an unsuccessful example where transfer learning was attempted from case study \#1 (source) to case study \#2 (target). 
Even in the $50$\%--$50$\% case, up-tilting BSs decreases the KPI in half of the dataset (the one from case study \#1). As a result, HD-BO fails to establish trust regions that include antenna up-tilts and gets trapped in an all-downtilt, low-performing local optimum.
\end{comment}

% Configurations-specific transfer learning
% On another note, we found that \emph{configuration-specific} transfer learning is also possible: using data collected from a certain scenario to optimize a new one where a particular antenna parameter has changed. The details are omitted due to space constraints. 
% The scenario source is based on GUEs and UAVs along corridors at 50\,m with BS antennas having a vHPBW of $30^{\circ}$, whereas the scenario targets involves BS antennas featuring a vHPBW of $10^{\circ}$.}
\section{Conclusion}
\label{sec:conclusion}

We employed data-driven optimization to jointly configure the BS antenna tilts and HPBWs of a real-world cellular network, achieving more than double the $10$\%-worst rates with respect to a 3GPP baseline. For scenarios involving UAVs, we identified configurations that improve their median rates fivefold, without degrading ground UE performance.
We further explored the transfer learning capabilities of our approach, using data from a scenario source to predict the optimal solution for a scenario target, reaching convergence with a similar number of iterations and negligible loss, without requiring a new initial dataset.

This work calls for multiple extensions. The parameters being optimized could include cell-specific beam codebook configurations or mobility management thresholds. The objective function maximization could be generalized to a multi-objective Pareto front: a set of non-dominated solutions where no objective can be improved without compromising another.

%\newpage
\bibliographystyle{IEEEtran}
\bibliography{journalAbbreviations, main}

\end{document}